\documentclass[10pt, conference]{IEEEtran}

\newcommand{\afblock}[1]{\noindent{\textbf{#1 }}}
\newcommand{\takeaway}[1]{\noindent{\textit{\textbf{Takeaway:}}} \textit{#1}}

\usepackage{times}
\usepackage[utf8]{inputenc}
\usepackage[T1]{fontenc}
\usepackage[nolist]{acronym}
\usepackage[colorlinks=true,allcolors=blue]{hyperref}
\usepackage{url}
\usepackage{pifont}
\usepackage{multirow}
\usepackage{siunitx}
\usepackage{array,booktabs}

\usepackage{tabularx}
\usepackage{cite}
\usepackage{amsmath,amssymb,amsfonts}
\usepackage{algorithmic}
\usepackage{graphicx}
\usepackage{textcomp}
\usepackage[usenames,dvipsnames,svgnames,table]{xcolor}
\usepackage{color,soul}
\usepackage{balance}
\usepackage{longtable}
\usepackage[caption=false]{subfig}
\usepackage[shortcuts]{extdash}
\usepackage{flushend} 

\usepackage{tikz}
\newcommand\copyrighttext{%
  \footnotesize \textcopyright~IFIP, 2023. This is the author's version of the work. It is posted here by permission of IFIP for your personal use. Not for redistribution. The definitive version was published in 2023 IFIP Networking Conference, \url{https://doi.org/10.23919/IFIPNetworking57963.2023.10186363}}
	\newcommand\copyrightnotice{%
	\begin{tikzpicture}[remember picture,overlay]
	\node[anchor=south,yshift=10pt] at (current page.south) {\fbox{\parbox{\dimexpr\textwidth-\fboxsep-\fboxrule\relax}{\copyrighttext}}};
	\end{tikzpicture}%
}

\begin{acronym}
    \acro{pep}[PEP]{Performance Enhancing Proxy}
    \acro{loops}[LOOPS]{Local Optimizations on Path Segments}
    \acro{rto}[RTO]{Retransmission timeout}
    \acro{fec}[FEC]{Forward Error Correction}
    \acro{holb}[HOL blocking]{Head-of-line blocking}
    \acro{quic}[QUIC]{Quick UDP Internet Connections}
    \acro{vpn}[VPN]{Virtual Private Network}
    \acro{masque}[MASQUE]{Multiplexed Application Substrate over QUIC Encryption}
    \acro{ietf}[IETF]{Internet Engineering Task Force}
    \acro{wg}[WG]{Working Group}
    \acro{tcpm}[TCPM]{TCP Maintenance and Minor Extensions}
    \acro{tcp}[TCP]{Transmission Control Protocol}
    \acro{bbr}[BBR]{Bottleneck Bandwidth and Round-trip propagation time}
    \acro{satcom}[SATCOM]{Satellite Communication}
    \acro{cca}[CCA]{Congestion Control Algorithm}
    \acro{iw}[IW]{Initial Window}
    \acro{alpn}[ALPN]{Application-Layer Protocol Negotiation}
    \acro{os}[OS]{Operating System}
    \acro{bdp}[BDP]{Bandwidth-Delay Product}
    \acro{rtt}[RTT]{Round-Trip Time}
    \acro{smaq}[SMAQ]{Secure Middlebox-Assisted QUIC}
    \acro{e2ee}[E2EE]{End-to-End Encrypted}
\end{acronym}

\acrodefplural{vpn}[VPNs]{Virtual Private Networks}
\acrodefplural{pep}[PEPs]{Performance Enhancing Proxies}
\acrodefplural{cca}[CCAs]{Congestion Control Algorithms}

\begin{document}

\title{Secure Middlebox-Assisted QUIC}

\author{Mike Kosek, Benedikt Spies, Jörg Ott\\
Technical University of Munich, Germany\\
\texttt{[kosek | spiesbe | ott]@in.tum.de}\\
\vspace{-2em}
}

\maketitle

\begin{abstract}
    While the evolution of the Internet was driven by the end-to-end model, it has been challenged by many flavors of middleboxes over the decades.
    Yet, the basic idea is still fundamental: reliability and security are usually realized end-to-end, where the strong trend towards ubiquitous traffic protection supports this notion.
    However, reasons to break up, or redefine the ends of, end-to-end connections have always been put forward in order to improve transport layer performance.
    Yet, the consolidation of the transport layer with the end-to-end security model as introduced by QUIC protects most protocol information from the network, thereby eliminating the ability to modify protocol exchanges.
In this paper, we enhance QUIC to selectively expose information to intermediaries, thereby enabling endpoints to consciously insert middleboxes into an end-to-end encrypted QUIC connection while preserving its privacy, integrity, and authenticity.
    We evaluate our design in a distributed Performance Enhancing Proxy environment over satellite networks, finding that the performance improvements are dependent on the path and application layer properties: the higher the round-trip time and loss, and the more data is transferred over a connection, the higher the benefits of \textit{Secure Middlebox-Assisted QUIC}.
    \copyrightnotice
\end{abstract}

{
  \let\thefootnote\relax\footnotetext{ISBN 978-3-903176-57-7\textcopyright~2023 IFIP}
}

\section{Introduction}
\label{sec:introduction}

The end-to-end model and networks doing just routing and
forwarding have served the evolution of the
Internet and a myriad of applications well.  Even though this principle
has been challenged by many flavors of middleboxes appearing over the
decades, it is still fundamental to service and content
delivery in the Internet: reliability, congestion control, and
security are usually realized by end-to-end (transport) connections.
The (recent) strong push towards ubiquitous traffic protection, naturally
end-to-end, emphasizes this.

Yet, reasons to break up---or redefine the ends of---end-to-end
connections have repeatedly been put forward, e.g., to improve 
performance for the user and/or the network operator.
Such optimizations may take different shapes, illustrated by:
\\
(a) Content Distribution Networks effectively ``cheat'' on the origin
server certificates to allow for faster content and service delivery
to the users from closer-by locations: they do maintain the end-to-end
transport but redefine the server-end~\cite{cdn-short-lived-tls}.
\\
(b) Operators of (sub)networks with path properties that are notably
different from the ``typical'' Internet characteristics often apply
flavors of connection splitting using \textit{Performance Enhancing Proxies}
(PEPs) to create independent control loops,
typically for congestion or error control, in order to speed up connections at
the transport layer~\cite{rfc3135}.
\\
(c) Live streaming contribution and distribution networks seek to push
media contents to a production system and then fan-out connections to
the consumers, effectively creating transport layer overlays. The
branching points in such overlays may need to perform rate adaptation
to match the capabilities of their downstream receivers but, at the
same time, shall not be able to access the content carried in those
streams~\cite{moq-work}.

The middleboxes in the above examples rely on
access to the information conveyed in the end-to-end connection and on
the ability to modify the protocol exchanges.
Because of this reliance, any such intermediate system has to make 
(implicit) assumptions about the end-to-end
protocol behavior.  Acting upon these assumptions may contribute to
the \textit{ossification} of the Internet as the \textit{expected}
behavior may become a prerequisite for traffic to pass now and in the
future~\cite{only-just-works,rfc7663}.  Thus, middleboxes---in particular the supposedly
\textit{transparent} ones---built with good intentions of performance
improvement, may hinder future network and protocol evolution.

There appears to be general consensus on protecting the end-to-end
information exchange from observation and modification inside the
network, rendering any sort of transparent middlebox a non-starter\footnote{
This consensus is witnessed, e.g., by the design of 
QUIC~\cite{rfc9000} to protect pretty much all protocol information 
from the network, or by over-the-top name resolution such as 
DNS-over-HTTPS~\cite{rfc8484} or DNS-over-TLS~\cite{rfc7858}.}.
This implies that introducing ``in-network'' functions like the above
require a conscious decision and consent by either or both endpoints
of an end-to-end connection to \textit{selectively expose information
  to specific nodes}.

Such controlled information exposure can basically happen in two
ways:
(1) \textit{In-band} of the end-to-end transport connection by explicitly
including middleboxes en route either during the initial setup as is
the case with explicitly chosen proxies, or by inserting them
later as could be achieved with redirection mechanisms, or
(2) \textit{Out-of-band} of the end-to-end transport connection by
establishing an independent signaling channel between one or both
endpoints and one or more middleboxes.
In both cases, the amount of information shared is controllable by the
endpoints: in the out-of-band case, this information is explicitly
compiled and sent to the middleboxes while, in the in-band case,
different levels of encryption can be used to selectively expose
information flowing end-to-end.

The intermediary functions themselves may be located \textit{on-path},
i.e., within the path determined by IP routing; this enabled
transparent middlebox operation in the past, e.g., if PEPs were on the default
route.  Or they may be \textit{off-path}, in which case they need to
be configured or actively discovered.  To achieve controlled
information exposure, endpoints need to become explicitly aware of and
consent to the middleboxes in the first place, which reduces the
former advantage of (transparent) on-path functions.

The actions a middlebox can sensibly perform depend on how much it is aware
of the protocol state and is authorized to change it, i.e., has the
appropriate keying material to interact with the endpoints.  Endpoints
may share detailed protocol state and thus enable modifying this
state at the cost of providing more insight into the application
interaction patterns (e.g., to adapt protocol behavior across different network
segments).
Without access to the protocol state, a middlebox does not become part
of the end-to-end protocol and thus the communication remains opaque,
which limits its operational capabilities to handling of the packets
passing through.  These may incur prioritizing, delaying, dropping,
marking, or otherwise shaping traffic~\cite{sidecar}, but also error
repair (retransmission, \textit{Forward Error Correction} (FEC))~\cite{tinyfecvpn}.  These may also
be applied to impact the protocol state indirectly, as the modified
packet flow (e.g., timing, losses) is interpreted by the protocol
state machines at the endpoints.
As a very recent example of an out-of-band signaling, 
Yuan et al.~\cite{sidecar} introduce a design that works with 
on-path middleboxes to perform packet scheduling
adaptation and also foresees other operations as a function of
information shared via explicit endpoint signaling.

\begin{figure}[t]
    \centering
\includegraphics[width=\linewidth,clip]{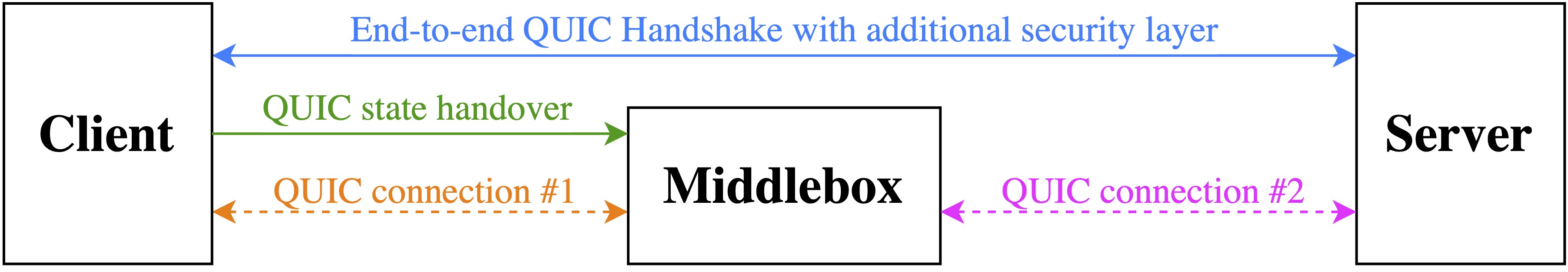}
	\caption{
		SMAQ design overview: Following an end-to-end QUIC \textit{Handshake} with an additional security layer (blue), the client hands over its QUIC state to a middlebox (green), which then splits the original end-to-end connection in-band into two independent connections using connection migration (orange and magenta).}
	\label{fig:hquic-pep-intro}
	\vspace{-1em}
\end{figure}

In this paper, we take a different road.  We explore an in-band design
to enable building middleboxes for QUIC in a way that preserves
the privacy, integrity, and authenticity of information end-to-end
while supporting QUIC-specific adaptation functions in selected
middleboxes.
Our design of \textit{Secure Middlebox-Assisted QUIC} (SMAQ, see Fig.\,\ref{fig:hquic-pep-intro} and §\ref{sec:methodology}) has three complementary elements: (1) a \textit{state
  handover mechanism} that allows endpoints to consciously insert a
  middlebox into an end-to-end encrypted QUIC connection and share keying material to operate on
selected protocol state; (2) \textit{enhanced QUIC connection
  migration} that enables directing connection traffic also to
off-path middleboxes; and (3) an \textit{additional security layer}
to preserve end-to-end security in spite of middleboxes.
The main non-functional goal is
leveraging readily available QUIC mechanisms such as connection migration 
and key exchange as much as possible.
We use the sample case of a distributed PEP 
to isolate the specifics of a satellite network segment as an evaluation scenario in
§\ref{sec:case-study}.  Following, we extensively discuss limitations in
§\ref{sec:limitations} before §\ref{sec:related-work} details related work
and we conclude in §\ref{sec:conclusion}. \section{Design}
\label{sec:methodology}

\textit{Secure Middlebox-Assisted QUIC} (SMAQ) allows endpoints to consciously insert middleboxes into an end-to-end encrypted 
QUIC connection while preserving its privacy, integrity, and authenticity.
We assume that the middlebox is trusted to a certain extent to perform
data forwarding and enhancement functions, e.g., since it is run by
the endpoint's network operator.  Note again that the degree of trust
is limited to QUIC protocol operation and \textit{excludes} access to
application data.
We stipulate further that trust into a middlebox implies
entitling it to add additional---mutually
trusted---middleboxes (of the same provider) as a practical
consideration since middleboxes might be in a much better position to
locate suitable further intermediaries for a given connection compared to the endpoint
having to discover those.

We focus on the design and initial evaluation of middlebox extensions
for QUIC connections and deliberately leave the discovery of
middleboxes as well as authorization and auditability for future
work (see §\ref{sec:limitations}).  In the following, we assume that a
client already discovered and established a connection to a middlebox
in the past
enabling 0-RTT connection establishment.
Moreover, we assume that QUIC \textit{Address Validation Using Retry Packets}~\cite[Sec. 8.1.2]{rfc9000} is disabled: While previous work showed that enforcing the \textit{traffic amplification limit} effectively safeguards against amplification attacks, \textit{Address Validation Using Retry Packets} can safely be skipped, thereby reducing first time connection establishments by 1$\times$RTT~\cite{quic.handshake}.

We first detail the SMAQ connection setup and its overhead in §\ref{subsec:hquic-handover} and §\ref{subsec:hquic-migration-time}, followed by §\ref{subsec:hquic-exception} outlining exception handling.
Subsequently, §\ref{subsec:hquic-state} highlights the state properties, and §\ref{subsec:hquic-xse} details the mechanisms leveraged to ensure end-to-end security of application data.
The section concludes with a discussion on the security considerations in §\ref{subsec:hquic-security}.

\subsection{Connection Setup}
\label{subsec:hquic-handover}

\begin{figure}[t]
	\centering
	\includegraphics[width=1.0\linewidth,trim=0 0 0 0, clip]{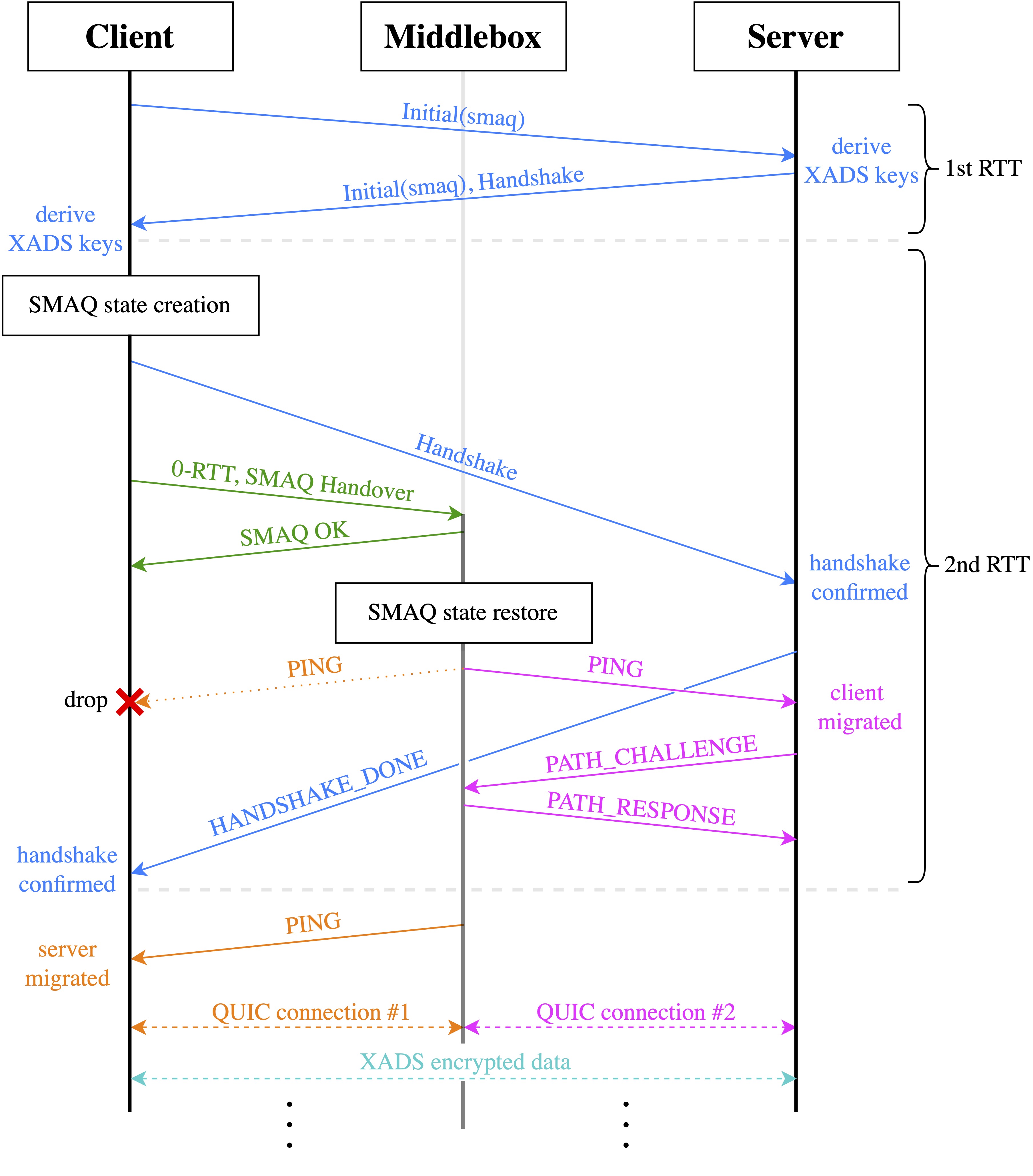}
	\vspace{-2em}
	\caption{SMAQ connection setup using out-of-band 0-RTT handover (green) following QUIC connection initiation (blue). Subsequent to the SMAQ state restore, QUIC \texttt{PING} frames are send to the client (orange) and server (magenta) to trigger QUIC connection migration on both endpoints, splitting the QUIC connection in-band into two independent control loops. Data is end-to-end encrypted between client and server with XADS keys (cyan).
	} \label{fig:hquic-handover}
	\vspace{-1em}
\end{figure}
 
The SMAQ connection setup is detailed in Fig.\,\ref{fig:hquic-handover}.
First, a QUIC connection between a client and a server is initiated (blue), where both indicate support for SMAQ by including the \texttt{smaq} transport parameter within the \textit{Initial} packet as defined by QUIC~\cite[Sec. 7.4]{rfc9000}.
Subsequently, both client and server derive keying material for an additional security layer on top of the QUIC transport security itself: The \textit{Extra Application Data Security} (XADS) is used for encrypting the application layer data, where the derived keys are maintained exclusively by client and server (see §\ref{subsec:hquic-xse}).
At this point, both the cryptographic keys for the QUIC connection itself and XADS are established.
Hence, SMAQ is able to parallelize the completion of the end-to-end QUIC connection (Fig.\,\ref{fig:hquic-handover} blue) and the state handover to the middlebox, where the latter is performed using an independently established\linebreak 0-RTT QUIC connection (Fig.\,\ref{fig:hquic-handover} green).
Because the SMAQ state contains the connection properties and cryptographic information of the original connection, excluding XADS keys (see §\ref{subsec:hquic-state}), the middlebox is able to restore the state in-band, splitting the connection into two independent control loops (Fig.\,\ref{fig:hquic-handover} orange and magenta).
This is achieved with QUIC's connection migration feature: Since every QUIC connection features a \textit{Connection ID} (CID), connections are identifiable independent of the endpoint IP addresses and port numbers.
Using non-zero-length CIDs, QUIC connections can be maintained even across IP address or port number changes~\cite[Sec. 5.1]{rfc9000}, e.g., when migrating to a new network.
Following the state restore, the middlebox sends QUIC \texttt{PING} frames to client and server, triggering connection migration on both endpoints while acting as a migrated server (client-facing, Fig.\,\ref{fig:hquic-handover} orange) and migrated client (server-facing, Fig.\,\ref{fig:hquic-handover} magenta), respectively.
Hence, SMAQ requires both the client and server endpoint to migrate (see §\,\ref{sec:limitations}).

The connection migration can only succeed once the handshake is \textit{confirmed}, i.e., the server has received the QUIC \textit{Handshake} packet, and the client has received the \texttt{HANDSHAKE\_DONE} frame~\cite[Sec. 9]{rfc9000} (Fig.\,\ref{fig:hquic-handover} blue).
Once the migration of the client completes (server-facing, Fig.\,\ref{fig:hquic-handover} magenta), the server initiates \textit{Path Validation} (\texttt{PATH\_CHALLENGE} and \texttt{PATH\_RESPONSE}) on the migrated address to verify the reachability~\cite[Sec. 8.2]{rfc9000}.
This validation can be skipped between the client and the middlebox as the reachability is already verified with the SMAQ state handover and its response (Fig.\,\ref{fig:hquic-handover} green).
Following connection migration, the connection is split into two independent control loops, and the middlebox splices the connection on behalf of the endpoints.
The end-to-end security of the application layer data is maintained (Fig.\,\ref{fig:hquic-handover} cyan) as they are protected by the XADS keys which are only known to the client and server (see §\ref{subsec:hquic-state}, §\ref{subsec:hquic-xse}).

Since QUIC connection migration is transparent to the application layer, the SMAQ handover is transparent as well.
However, it must be prevented that client and server directly exchange data during the handover process to ensure that the restored state on the middlebox is consistent with the handed over state.
Since the client is the initiator of the SMAQ handover, it simply does not send any data until the SMAQ connection setup is completed.
The server, on the other hand, does not have any knowledge if a handover will be performed, and therefore may send data; hence, the client is required to drop any received data from the original server.

\subsection{Connection Setup Overhead}
\label{subsec:hquic-migration-time}

A client of a regular QUIC connection is able to send application data once the required keys are established, which corresponds to 1$\times$RTT.
Using SMAQ, however, the client cannot send application data until the migration is completed, i.e., the client has received the \texttt{PING} following the \texttt{HANDSHAKE\_DONE} frame (Fig.\,\ref{fig:hquic-handover} orange).
While the \texttt{HANDSHAKE\_DONE} is received after 2$\times$RTTs, the time required for the \texttt{PING} to arrive depends on the one-way delay between middlebox and client and on the retransmission timer of the middlebox.
Hence, in the best case, the SMAQ connection setup requires slightly more than 2$\times$RTTs if the middlebox is located on-path and close to the client (e.g., within the same local network), resulting in an initial overhead of SMAQ compared to QUIC of slightly more than 1$\times$RTT.
At this point, our design does not consider 0-RTT connection establishment between client and server, which may be exploited to further optimize SMAQ connection establishment.

\subsection{Exception Handling}
\label{subsec:hquic-exception}
§\ref{subsec:hquic-handover} and §\ref{subsec:hquic-migration-time} describe a successful connection setup, but various issues may arise:
If one endpoint does not (wish to) support SMAQ, the \texttt{smaq} transport parameter is ignored~\cite[Sec. 7.4]{rfc9000} and a regular QUIC connection is established: a SMAQ middlebox cannot be added to the connection\linebreak unilaterally.

Moreover, the server typically receives the QUIC \texttt{PING} after the QUIC \textit{Handshake}, at which point the \texttt{HANDSHAKE\_DONE} frame was already sent to the client.
However, the \texttt{HANDSHAKE\_DONE} frame may be retransmitted after the migration of server and middlebox already completed.
In this case, the frame is send to the middlebox, which forwards it to the client for the handshake to succeed.
Furthermore, an endpoint may receive the middlebox \texttt{PING} before the connection is confirmed, which is also the expected behavior on the client.
In compliance with QUIC~\cite[Sec. 9]{rfc9000}, the endpoints do not update the path on those early frames from another address, but the middlebox resends it on its retransmission timer until the migration completes or the connection migration fails (Fig.\,\ref{fig:hquic-handover} red mark, illustrated for client only).
If the migration fails for either endpoint, the whole connection is closed by an error or times out.
Recovery mechanisms for failed migrations is subject of future work.

\begin{table}[t]
    \centering
    \caption{SMAQ state properties.}
    \definecolor{rowgray}{gray}{0.9}
\centering
    \begin{tabular}{lll}
        \toprule
            \textbf{Parameter} & \textbf{Description} \\
        \midrule
            Active Connection IDs & Active Connection IDs of client and server \\ 
            & with the associated sequence numbers \\
            \rowcolor{rowgray}
            Stateless reset tokens & All tokens with the associated Connection IDs \\
            QUIC version & Used QUIC version \\
            \rowcolor{rowgray}
            Cipher suite & Used TLS cipher suite \\
            Key phase & Number of the current key phase \\
            \rowcolor{rowgray}
            Current traffic secrets & Client and server phase traffic secrets, i.e., \\
            \rowcolor{rowgray}
            & \textit{<sender>\_application\_traffic\_secret\_<phase>} \\
            Header protection keys & Client and server header protection keys, i.e., \\
            & \textit{<sender>\_header\_protection\_key}\\
            \rowcolor{rowgray}
            Endpoint addresses & IP addresses and ports of client and server \\
            Transport parameters & Sent transport parameters of client and server \\
            \rowcolor{rowgray}
            Packet numbers & Highest sent and received packet numbers \\
\bottomrule
    \end{tabular}
    \label{tab:hquic-state}
\end{table}

\subsection{State Properties}
\label{subsec:hquic-state}

For state handover, the client creates a concise, serialized state object containing only the essential connection properties and cryptographic information of the original connection (excluding XADS keying material) required to restore its state on the middlebox (see Tab.\,\ref{tab:hquic-state}).
The state object can be created as soon as the cryptographic keys for the QUIC connection are established, i.e., the client received the QUIC \textit{Handshake} packet~\cite{rfc9001}.
The state can only be restored if the QUIC implementation of the middlebox supports the QUIC version, cipher suite, and all transport parameters that are handed over.
If at least one requirement fails, the middlebox rejects the handover with a \texttt{SMAQ Error} message, and the regular QUIC connection is continued.
If all requirements are met, the successful handover is acknowledged with \texttt{SMAQ OK} (see Fig.\,\ref{fig:hquic-handover}, green).

\subsection{Extra Application Data Security (XADS)}
\label{subsec:hquic-xse}
To ensure end-to-end security of application data between client and server, we present \textit{Extra Application Data Security} (XADS) which provides an additional security layer on top of QUIC.
Since QUIC without XADS uses the same cryptographic keys to protect transport and application data, access to one security context cannot be shared with a middlebox independently from the other.

\begin{figure}[t]
	\centering
	\includegraphics[width=1.0\linewidth,trim=0 0 0 0, clip]{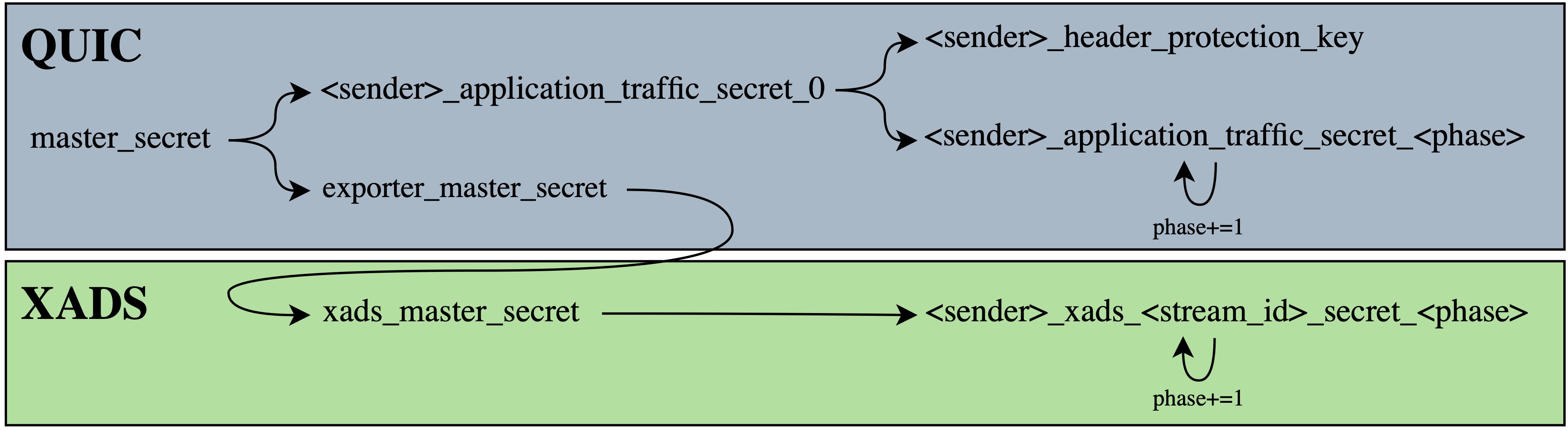}
	\caption{Derivation of cryptographic keys for QUIC and XADS. Due to forward secrecy provided by HKDF, keys can only be derived in arrow direction. \textit{<sender>} is either client or server, \textit{<stream\_id>} is the ID of the stream to protect. \textit{<phase>} is the key phase, incremented by key updates.
	} \label{fig:hquic-key-derivation}
\end{figure}

The XADS keying material remains on the endpoints, i.e, it is not shared with the middlebox (see §\ref{subsec:hquic-state}).
Moreover, it relies on the cryptographic TLS 1.3 handshake incorporated by the QUIC connection establishment which is detailed in Fig.\,\ref{fig:hquic-key-derivation}.
After having received the \textit{Initial} packet, both client and server derive the \texttt{xads\_master\_secret} from the \texttt{exporter\_master\_secret} using TLS keying material exporters~\cite[Sec. 7.5]{rfc8446}~\cite{rfc5705}.
The key derivation is a one-way pseudorandom function; i.e., the \texttt{exporter\_master\_secret} can be used to derive arbitrary XADS key material, but not vice versa.
XADS uses the TLS 1.3 record protocol over QUIC streams.
For every opened unidirectional QUIC stream a new secret is derived from the \texttt{xads\_master\_secret}.
XADS encapsulates the application data into TLS records, which are protected by the corresponding client or server secret of the current key phase; e.g., \texttt{client\_xse\_0\_secret\_1} for the client's stream ID 0 within the 1st key phase.
Hence, by deriving the XADS keys from the cryptographic TLS 1.3 handshake incorporated by QUICs connection establishment, no additional handshakes, and therefore no additional round trips, are required for XADS.

After the first secret (i.e., key phase 0) for the XADS stream has been derived from the \texttt{xads\_master\_secret}, it is cryptographically independent of other streams, as well as the other direction of a bidirectional stream.
Hence, the traffic secrets of every stream and direction can be updated independently leveraging TLS 1.3 \texttt{KeyUpdate}~\cite[Sec. 4.6.3]{rfc8446}:
When the lifetime of a traffic key is reached, a new key is generated from the key of the previous phase, where the forward secrecy relies on the Expand-Label function of HKDF (HMAC-Based Key Derivation Function,~\cite{rfc5869}).

While TLS records are of variable length, a record can carry at most 2\textsuperscript{14} bytes of data with a minimum overhead of 22 byte per record~\cite[Sec. 5.2]{rfc8446}; hence, the overhead induced by XADSs' TLS record encapsulation is at least $\sim$0.13\,\%.
Additionally, we investigated possible performance penalties of XADS in comparison to regular QUIC, where we did not find any significant differences in our scenarios while using Hardware-assisted \texttt{AES} ciphers; a systematic performance evaluation is left for future work.

\subsection{Security Considerations}
\label{subsec:hquic-security}

SMAQ enables endpoints to consciously insert middleboxes into an end-to-end encrypted QUIC connection while preserving its privacy, integrity, and authenticity.
The fundamental prerequisite for this is a secure end-to-end key exchange.
With SMAQ, the QUIC handshake remains end-to-end, enabling to securely exchange the SMAQ transport parameter and derive the XADS keys directly between endpoints~\cite[Sec. 7.4]{rfc9000}.

A SMAQ state contains the header protection keys and the traffic secrets of the current key phase (see Tab.\,\ref{tab:hquic-state}).
Since the key derivation function is considered one-way~\cite[Sec. E.2]{rfc8446}, traffic secrets of previous key phases, and secrets such as the exporter master secret, are not exposed.
However, with the information contained in the SMAQ state, a middlebox has full access to the QUIC connection itself, excluding the XADS protected application data.
Therefore, the SMAQ state must be protected from access by third parties, must only be transmitted on encrypted and authenticated channels, and should be erased as soon as the state is no longer required.

Although all application data remains end-to-end encrypted, the middlebox can infer information of the application layer using metadata, e.g., by observing length and timing of encrypted records as discussed in~\cite[Sec. E.3]{rfc8446}.
Moreover, a middlebox can also analyze individual stream behavior, which can reveal information about different application contexts as they are likely carried on different streams.
Furthermore, a middlebox can also manipulate, drop, or inject, frames, which could cause unexpected application layer behavior.

Exposing some information to the middlebox is a necessary tradeoff for its capabilities.
Hence, a minimum level of trust is required between clients and middleboxes.
While our work focusses on the design and evaluation of SMAQ, we leave authorization, accountability, and auditability open for future work (see §\ref{sec:limitations}).

\takeaway{SMAQ enables endpoints to consciously insert middleboxes into an end-to-end encrypted QUIC connection while preserving its privacy, integrity, and authenticity: the connection state of an endpoint is handed over to a middlebox, thereby  splitting the connection in-band into two independent control loops using connection migration.
Yet, the end-to-end security is ensured with an additional encryption layer on top of QUIC's encryption.
}

 \section{Case Study: Distributed Performance Enhancing Proxies}
\label{sec:case-study}

We now apply SMAQ to realize \textit{Performance Enhancing Proxies} (PEPs) for QUIC connections, a typical use case for splitting end-to-end connections into multiple independent control loops~\cite{rfc3135,pepsal_paper,qpep,quic.satcom,jones-tsvwg-transport-for-satellite-02}.
\textit{Distributed} PEPs placed on an ingress and an egress point of a network can be used to enhance the transport connection within the enclosed path segment by applying path specific optimizations.
While the design presented in §\ref{sec:methodology} illustrates the use of a single middlebox, SMAQ supports transitive state handover to multiple middleboxes, thereby enabling a distributed PEP setup.

Fig.\,\ref{fig:hquic-handover-pep} shows the simplified SMAQ connection setup using two distributed PEPs.
Following the QUIC connection initiation (blue), the state is handed over out-of-band from the client to PEP\,\#1 (green).
Subsequently, a new altered SMAQ state is created by PEP\,\#1, where the client address is replaced with the address of PEP\,\#1 itself.
This new state is send to PEP\,\#2 out-of-band (light green), which then migrates the connection in-band to PEP\,\#1 (magenta), as well as to the server (petrol); subsequently, the server initiates \textit{Path Validation} on the migrated address in order to verify the reachability.
This validation is not required on the path between client and PEP\,\#1, as well as the path between PEP\,\#1 and PEP\,\#2, as the reachability is already verified with the SMAQ state handover.
The end-to-end path now consists of three individual QUIC connections, splitting the connection into three independent control loops.
Yet, the end-to-end security of application layer data between the endpoints is maintained using XADS (cyan).

While the distributed PEP setup includes two handovers, the initial overhead in terms of required round trips (see Fig.\,\ref{fig:hquic-handover-pep}, rightmost) in comparison to QUIC is identical to a single handover, i.e., slightly more than 1$\times$RTT in the best case if both PEPs are on-path and PEP\,\#1 is close to the client (e.g., within the same local network).
In this setup, the distributed PEPs can enhance the transport connection on the enclosed path segment, e.g., by adjusting QUIC parameters like the \textit{Congestion Control Algorithm} (CCA) or the \textit{Initial Window}.

\subsection{Test Environment}
\label{subsec:environment}

\begin{figure}[t]
	\centering
	\includegraphics[width=1.0\linewidth,trim=0 0 0 0, clip]{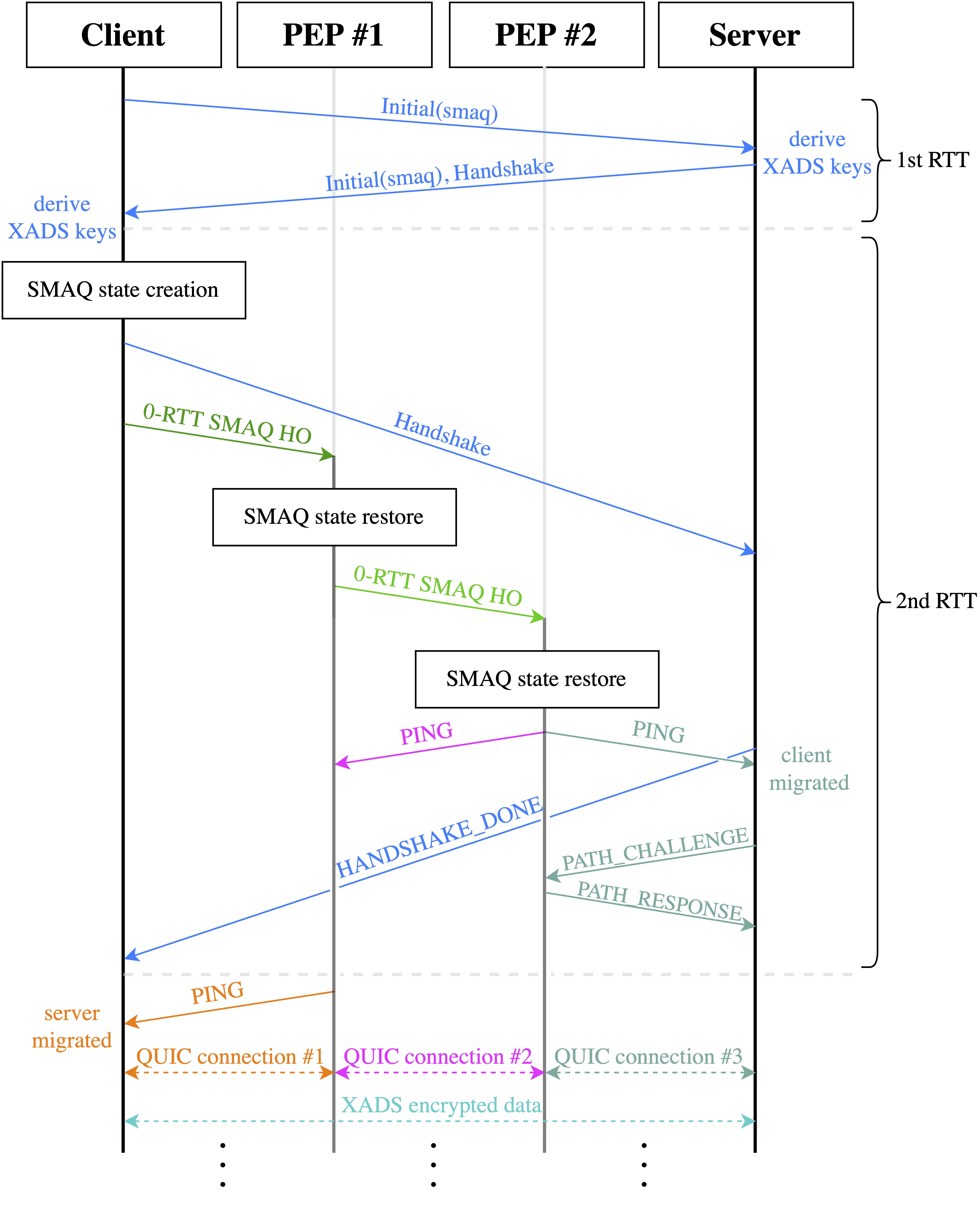}
	\caption{Simplified SMAQ connection setup using out-of-band 0-RTT handover (green, light green) following QUIC connection initiation (blue) using 2 distributed PEPs, splitting the QUIC connection in-band into 3 independent control loops (orange, magenta, petrol). Data is end-to-end encrypted between client and server with XADS keys (cyan).
	} \label{fig:hquic-handover-pep}
	\vspace{-1em}
\end{figure}
 
For our case-study, we implement SMAQ by extending \textit{quic-go}, building \texttt{smaq\=/pep} to provide PEP optimizations, \texttt{smaq\=/perf} to perform \textit{Middlebox Migration Time} and \textit{Bulk Download} measurements, as well as \texttt{smaq\=/http\=/perf} to perform \textit{Web Peformance} measurements in a distributed PEP environment using the open source \textit{Satellite Communication Emulation Testbed}~\cite{quic.satcom}.
The testbed enables reproducible measurements over SATCOM networks while featuring link-layer emulation using \textit{OpenSAND}~\cite{opensand}.
Thereby, the testbed follows the distributed PEP setup as presented in Fig.\,\ref{fig:hquic-handover-pep}, where the PEPs are placed on the ingress and egress point of the SATCOM network in order to optimize the transport connection in between.
To enable the reproduction of our findings, we make the developed tools publicly available, aiming to facilitate future studies using SMAQ~\cite{smaq}.

Using the default settings of the \textit{Satellite Communication Emulation Testbed}, the link-layer goodput in server to client direction is parametrized with 20\,Mbps, and two satellite orbits (\textit{Low Earth Orbit} (LEO) and \textit{Geostationary Orbit} (GEO)) with two loss profiles (random distribution of 0.01 and 0.1\,\% loss) are evaluated.
While 0.01\,\% loss represents real world satellite conditions, 0.1\,\% loss is considered an edge case~\cite{quic.satcom}.
PEP\,\#1 ist placed within the local network of the client; hence, the RTT between client and PEP\,\#1 is below 1\,ms.
Moreover, we optimize the retransmission timer of the client-facing connection of PEP\,\#1 in order to reduce the overhead of the time required for the retransmitted \texttt{PING} to arrive at the client following the \texttt{HANDSHAKE\_DONE} (see §\ref{subsec:hquic-handover}), thus optimizing the SMAQ connection setup: the \texttt{Initial RTT} estimation is set to the \texttt{Smoothed RTT} from a previous connection~\cite[Sec. 5]{rfc9002}, and exponential backoff is disabled until the migration succeeds.
For the GEO orbit, the one-way delay of the SATCOM connection between PEP\,\#1 and PEP\,\#2 is set to 250\,ms as derived by the speed of light in a vacuum, where the LEO one-way delay is set to 16\,ms based on measurements performed using Starlink~\cite{quic.satcom}.
Moreover, the one-way delay between PEP\,\#2 and the server is configured with 40\,ms for both GEO and LEO orbits in order to emulate their terrestrial distance, resulting in a total RTT of 580\,ms for GEO, and 112\,ms for LEO~\cite{quic.satcom}.
While both PEPs are placed on-path, and PEP\,\#1 is located within the local network of the client, our case study represents a best case environment for SMAQ where the initial overhead corresponds to slightly more than 1$\times$RTT in comparison to QUIC.

Every combination of satellite orbit and loss profile is measured using an end-to-end QUIC connection (dubbed \textit{QUIC}), as well as a PEP-optimized SMAQ connection (dubbed \textit{SMAQ\=/PEP}), resulting in a total of 8 measurement scenarios.
The end-to-end \textit{QUIC} measurements use the default QUIC \textit{Congestion Control Algorithm} (CCA) based on \textit{NewReno} with an initial congestion window of 10 packets~\cite[Sec. 7]{rfc9002}.
For \textit{SMAQ\=/PEP}, we use the identical settings for both client and server, but optimize the SATCOM transport connection with the distributed PEPs by using \textit{Hybla-Westwood}~\cite{hybla-westwood} as CCA:
While \textit{Hybla}~\cite{hybla} improves the congestion window increase on high latency connections by being more aggressive in comparison to \textit{NewReno}, \textit{Westwood}~\cite{westwood} improves the goodput over links with high packet loss by continuously estimating the usable bandwidth in order to minimize the congestion window reduction on non-congestion induced packet loss.

\subsection{Evaluation}
\label{subsec:evaluation}

We begin our evaluation by analyzing the \textit{Middlebox Migration Time} of SMAQ\=/PEP, followed by \textit{Bulk Download} measurements for both SMAQ\=/PEP and QUIC connections.
We then present \textit{Web Performance} measurements, highlighting the potential benefits of SMAQ\=/PEP in comparison to QUIC for web browsing.
All measurements are performed using QUIC version 1, where the \textit{Web Performance} measurements leverage HTTP/3~\cite{rfc9114} on top of QUIC.

\afblock{Middlebox Migration Time.}
We first verify our assumptions of the initial overhead of the SMAQ\=/PEP connection setup in comparison to QUIC. For this, we measure the time between the client creating the SMAQ state (corresponding to the client receiving the \textit{QUIC Handshake} packet) until SMAQ\=/PEP is able to send application data (corresponding to the client receiving \texttt{PING}, see Fig.\,\ref{fig:hquic-handover-pep} and §\ref{subsec:hquic-migration-time}).
Since our case study represents a best case environment for SMAQ\=/PEP (see §\ref{subsec:environment}), the \textit{Middlebox Migration Time} should correspond to slightly more than 1$\times$RTT.
The measurements are repeated 100 times for both GEO and LEO orbits for 0.01 and 0.1\,\% loss. For the \textit{Middlebox Migration Time}, we find a median of $\sim$585\,ms for GEO, and a median of $\sim$117\,ms for LEO, each for both loss scenarios.
Comparing the observations to the expected RTTs of 580\,ms for GEO and 112\,ms for LEO, we find a difference of $\sim$5\,ms which we attribute to the time required for the creation of the states and their restoration, and the time required for the \texttt{PING} to arrive at the client following the \texttt{HANDSHAKE\_DONE} (see §\ref{subsec:hquic-migration-time}).
Hence, the results confirm our assumptions of the initial overhead of the SMAQ\=/PEP connection setup to be slightly above one end-to-end RTT.

\begin{figure}[t]
    \centering
\includegraphics[width=1\linewidth,trim=7 7 7 0,clip]{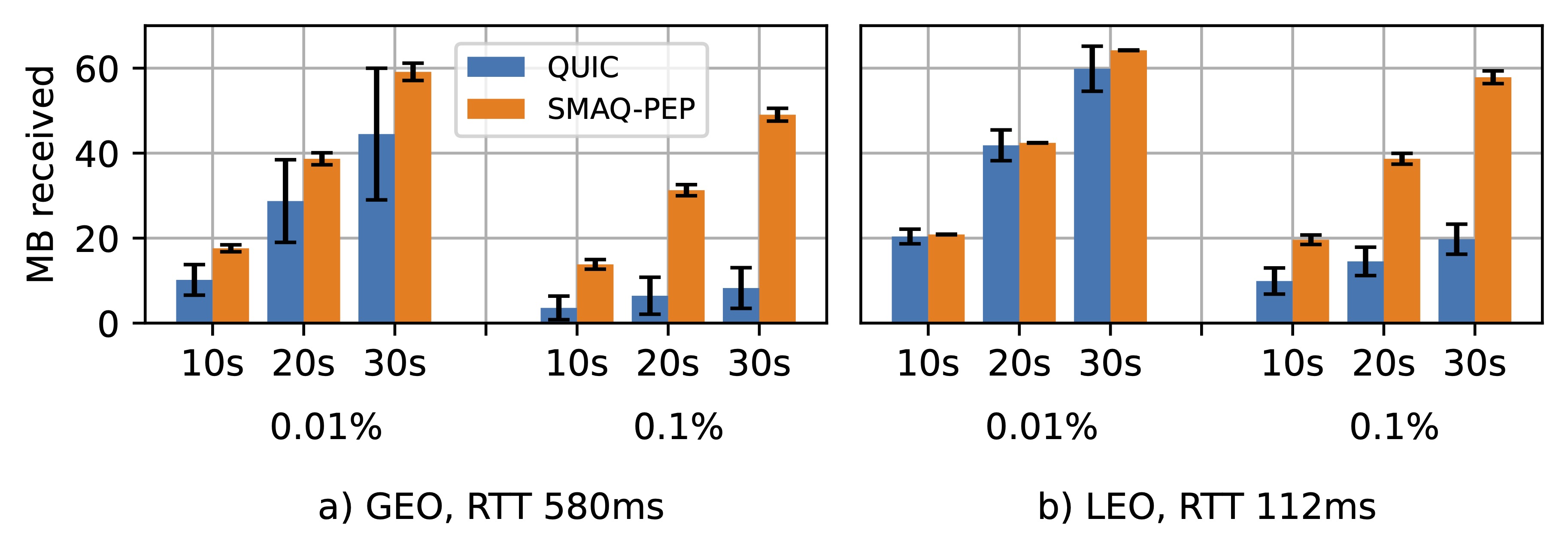}
\caption{Median received bytes after 10, 20, and 30 seconds bulk download for 0.01 and 0.1\,\% loss in GEO (a, left) and LEO (b, right) orbits using QUIC (blue) and SMAQ\=/PEP (orange).}
	\label{fig:goodput}
\end{figure}

\afblock{Bulk Download.}
For \textit{Bulk Download}, we evaluate the bytes received by the client over multiple time intervals:
Following connection establishment, the client sends an application layer request to the server, which in turn sends randomized data to the client while maximizing its goodput.
Fig.\,\ref{fig:goodput} presents the received bytes after 10, 20, and 30 seconds bulk download following the client's QUIC \textit{Initial} for 0.01 and 0.1\,\% loss in GEO (a, left) and LEO (b, right) orbits using QUIC (blue) and SMAQ\=/PEP (orange).
The measurements are repeated 100 times per scenario, and we present the medians as well as the standard deviations over all measurement runs.

Evaluating the GEO orbit with an RTT of 580\,ms (a, left), we find that the client receives more bytes using SMAQ\=/PEP in comparison to QUIC in every time interval and for every loss configuration despite the initial overhead of the \textit{Middlebox Migration}.
Analyzing 0.01\,\% loss, we find a maximum relative increase of SMAQ\=/PEP in comparison to QUIC with $\sim$73\,\% after 10\,s, decreasing to $\sim$33\,\% after 30\,s.
Yet, the standard deviation of SMAQ\=/PEP with $\sim$1--2\,MB is lower in comparison to QUICs $\sim$4--15\,MB.
Evaluating 0.1\,\% loss, we observe the same trends; however, the benefit of SMAQ\=/PEP in comparison to QUIC is more pronounced with a relative increase of bytes received with up to $\sim$494\,\% after 30\,s.

For the LEO orbit with an RTT of 112\,ms (b, right), we observe that the received bytes using SMAQ\=/PEP are comparable to QUIC for 0.01\,\% loss, where the standard deviation is again lower using SMAQ\=/PEP.
Yet, despite the initial overhead of the \textit{Middlebox Migration}, SMAQ\=/PEP does increase the bytes received in comparison to QUIC after 10 and 20\,s slightly with $\sim$1--2\,\%, increasing to $\sim$7\,\% after 30\,s.
Evaluating 0.1\,\% loss, we again observe a strong benefit of SMAQ\=/PEP in comparison to QUIC with a maximum relative increase of $\sim$193\,\% at 30\,s, accompanied by a lower standard deviation of $\sim$1--2\,MB in comparison to QUICs $\sim$3--4\,MB.

Combining our observations for both GEO and LEO orbits, we find that the benefits of SMAQ\=/PEP increase the more loss is present, resulting in an increase in bytes received with a lower standard deviation despite the initial overhead of the \textit{Middlebox Migration}.
We attribute the benefits to the usage of the \textit{Hybla-Westwood} CCA on the SATCOM connection (see §\ref{subsec:environment}):
With \textit{Westwoods} resiliency against packet loss, and the advantages of \textit{Hybla} on high latency connections, the break-even of SMAQ\=/PEP in comparison to QUIC is reached $\sim$1.9\,s ($\sim$3.3$\times$RTTs) on GEO orbits, respective $\sim$0.6\,s ($\sim$5.4$\times$RTTs) on LEO orbits following the client's QUIC \textit{Initial}.

\begin{figure}[t]
    \centering
	\includegraphics[width=\linewidth]{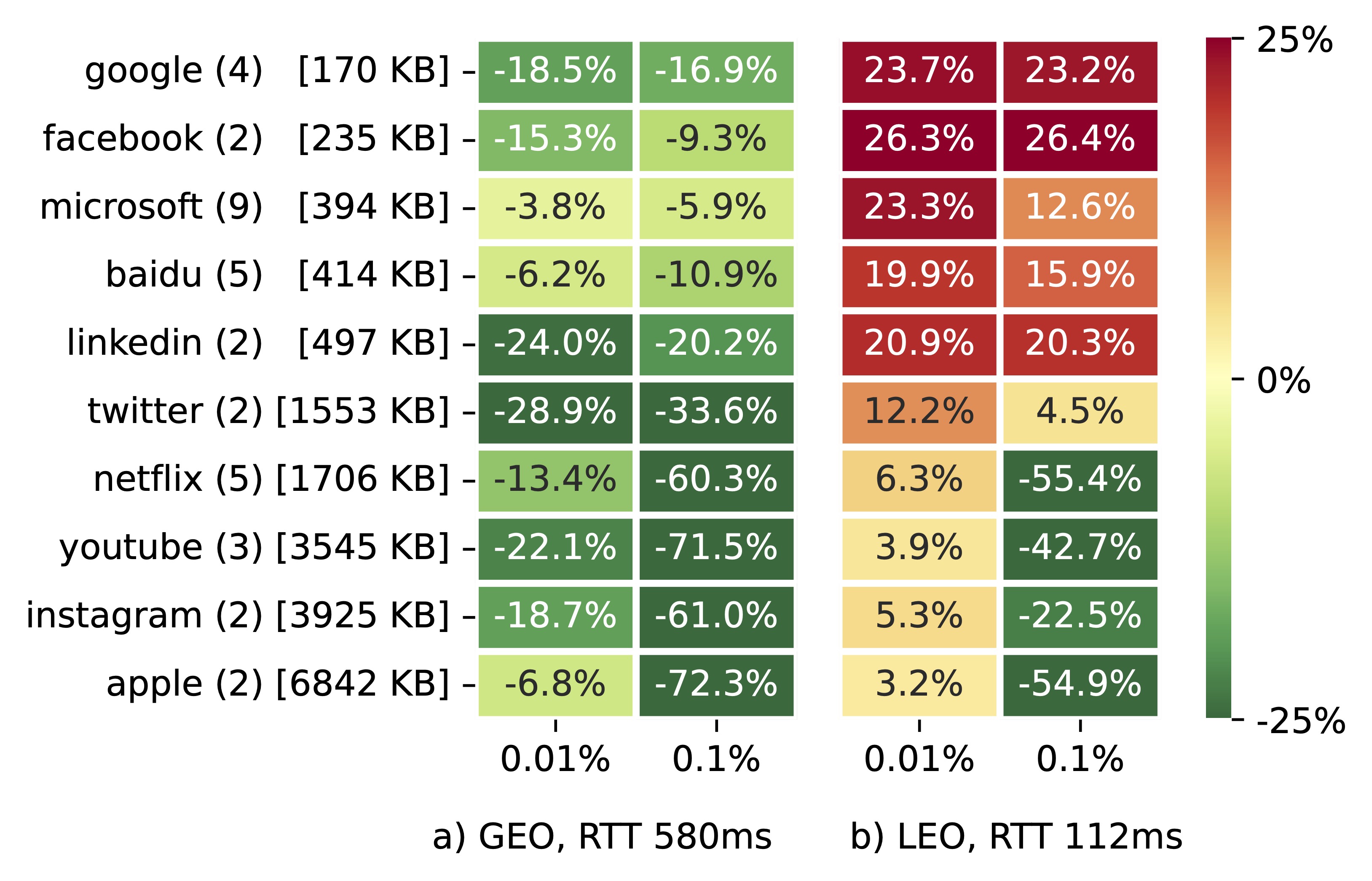}
\caption{Relative median difference of approximated Page Load Time (aPLT) of SMAQ\=/PEP in comparison to QUIC for 0.01 and 0.1\,\% loss in GEO (a, left) and LEO (b, right) orbits for Tranco top 10 webpages. Number of connections established by each webpage are in parenthesis. Sorted ascending by the average number of bytes transferred per connection (in square brackets). Negative values colored in green indicate a faster page load using SMAQ\=/PEP.}
\label{fig:plt}
\end{figure}

\afblock{Web Peformance.}
For our \textit{Web Performance} measurements, we evaluate the \textit{Page Load Time} using HTTP/3 over SMAQ\=/PEP in comparison to HTTP/3 over QUIC for the top 10 most popular webpages from the research-oriented \textit{Tranco} top list~\cite{tranco} as of December 22, 2022.
We first download the webpages leveraging \texttt{wget} with a User-Agent representing Chrome 107, ensuring that all elements from all hostnames required to render the webpage (images, fonts, scripts, etc.) are downloaded.
For the sake of simplicity, JavaScripts are not executed, i.e., resources that are requested by those are not considered.
Subsequently, we issue self-signed TLS certificates for each hostname of every webpage in order to serve all hostnames within our emulation testbed on dedicated servers, requiring the client to establish a new connection for each hostname requested; with this, a realistic HTTP/3 client behavior is obtained.
Therefore, the client requests all elements from all hostnames required to render the webpage, enabling the approximation of the \textit{Page Load Time}: Since SMAQ is not yet implemented in browsers, we evaluate the \textit{approximated Page Load Time} (aPLT) by leveraging \texttt{smaq-perf}, measuring the time between the client's QUIC \textit{Initial} until all required elements are received.
Since both SMAQ\=/PEP and QUIC are measured with the outlined methodology, our comparative evaluation of the relative differences of the aPLT enables us to assess the potential benefits of SMAQ\=/PEP in comparison to QUIC in a typical web browsing use-case.

Fig.\,\ref{fig:plt} presents the median relative aPLT difference of SMAQ\=/PEP in comparison to QUIC for 0.01 and 0.1\,\% loss in GEO (a, left) and LEO (b, right) orbits, where the measurement are repeated 100 times per scenario.
The webpages are sorted ascending from top to bottom by the average number of bytes transferred per connection (in square brackets), where the number of connections established by each webpage are presented in parenthesis.
E.g., while 4 connections are established by requesting \textit{google}, 170\,KB are transferred over each of the 4 connections on average.

Evaluating the GEO orbit (a, left), we find that the page load using SMAQ\=/PEP improves over QUIC for every webpage and loss configuration with up to $\sim$29\,\% for 0.01\,\% loss (\textit{twitter}), and up to $\sim$72\,\% for 0.1\,\% loss (\textit{apple}).
On the other hand, the LEO orbit (b, right) shows that SMAQ\=/PEP does prolong the page load in comparison to QUIC for every webpage for the 0.01\,\% loss scenario, where for 0.1\,\% loss SMAQ\=/PEP does improve over QUIC in 4 out of the 10 webpages.

Combining our observations for both GEO and LEO orbits, we again find that the benefits of SMAQ\=/PEP increase the more loss is present; yet, the initial overhead of the \textit{Middlebox Migration} does prolong the page load for most webpage/loss configurations on the faster LEO satellite connection.
Next to the orbit and loss configuration, we find that the architecture of the webpages is the most decisive factor for the observed relative differences.
Comparing the average number of bytes transferred per connection for each webpage (Fig.\,\ref{fig:plt}, square brackets, sorted ascending), we find that SMAQ\=/PEP tends to be more beneficial in comparison to QUIC the more bytes are transferred per connection.
Hence, we constitute that the higher the RTT and loss, and the more data is transferred over a connection, the likelier that the initial overhead of the \textit{Middlebox Migration} is overcome, which in turn leads to a higher benefit of SMAQ\=/PEP.
However, the overhead is typically only induced once per hostname: when browsing the website, requests re-use the already established connections, and subsequent pageloads directly benefit from SMAQ\=/PEP.

\takeaway{Our case-study shows the potential benefits of SMAQ applied in a distributed PEP environment: While the initial overhead of the \textit{Middlebox Migration} does add $\sim$1\,RTT to the connection setup, the break-even is reached after $\sim$1.9\,s ($\sim$3.3\,RTT) on GEO orbits and $\sim$0.6\,s ($\sim$5.4\,RTT) on LEO orbits for \textit{Bulk Downloads}.
While the \textit{Page Load Time} improves over GEO orbits using SMAQ, however, most LEO page loads are prolonged, thereby showing the dependency to the path properties and webpage architecture: The higher the RTT and loss, and the more data is transferred over a connection, the higher the potential benefits of SMAQ.
}
 \section{Limitations and Future Work}
\label{sec:limitations}
 
While SMAQ enables endpoints to consciously insert middleboxes into an end-to-end encrypted QUIC connection while preserving its privacy, integrity, and authenticity, its design is currently in an early stage with multiple open challenges.

\afblock{Authorization, Accountability and Auditability.}
With SMAQ, middleboxes have full access to control information of QUIC connections.
Hence, middleboxes can manipulate, drop, or inject, frames, or even migrate the connection to additional middleboxes without being accountable for any changes made.
Our design therefore requires a minimum level of trust which can be achieved on the basis of, e.g., a secure credential exchange, or \textit{Public Key Infrastructure} (PKI).
Moreover, we are exploring more fine-grained controls with a least-privilege approach, i.e., restricting the access of middleboxes to information which are required for its task, allowing access only to specific frame types for example.
For this purpose, multiple encryption contexts could be leveraged as proposed by Naylor et al.~\cite{mctls}.
Further, the mechanisms introduced by Lee et al.~\cite{matls} could be adopted in order to provide accountability and auditability.

\afblock{Server Migration.}
Our design requires both the client and server endpoint to migrate, where server migration was omitted from QUIC version 1 in order to reduce its complexity~\cite{quic-mailinglist-server-migration}.
While our work shows a general application for server migration, it is not unprecedented: interest around the concept sparks in the area of container migration at the edge~\cite{quic-container-migration,etsi-mec-migration}, justifying an exploration for future QUIC versions.

\afblock{NAT Traversal.}
Our design assumes that all addresses between handover-partners are reachable, which is sufficient for constellations where the handover does not cross network domains (e.g., distributed PEPs where all addresses are reachable within a network segment, see §\ref{sec:case-study}).
Yet, \textit{Network Address Translation} (NAT) traversal must be considered, where two challenges arise:
1)~The client might not know its public address assigned by the NAT and therefore cannot pass it on to the middlebox, and 2), the NAT may not allow ingress traffic from an unknown address, e.g., the middlebox.
A solution for this is to multiplex the out-of-band SMAQ state handover and the migrated in-band QUIC connection over the same UDP port:
While the handover is initiated by the client, a NAT binding is created on the gateway between client and middlebox, where the same IP address and port are subsequently reused for the handover of the QUIC connection.
However, an open problem still to overcome is the possible collision of \textit{Connection IDs}.

\afblock{Handovers and Connection Establishment.}
The presented work does currently only consider client-initiated handovers; however, the design does also allow for server-initiated handovers which we are currently exploring.
Moreover, handovers are currently limited to be performed during the QUIC handshake in order to simplify state creation and restoration.
Yet, endpoints can make more informed decisions about potential benefits of an SMAQ handover following connection establishment, i.e., based on transport and/or application layer observations (e.g., experienced RTT, requested HTTP/3 payload).
We will therefore investigate handovers which are performed at arbitrary times during the lifetime of a connection in the future.
In addition, the interplay of 0-RTT client-server handshakes and SMAQ is not yet considered, showing potential to further optimize SMAQ connection establishment.

\afblock{Feature Negotiation.}
SMAQ requires all middleboxes to support the QUIC features negotiated between client an server, e.g., version, cipher suite, or extensions.
Clients could be informed about supported features of middleboxes during discovery and authentication; hence, they can limit their offered features within the QUIC \textit{Initial} to an intersection between client and middlebox capabilities.
However, this hinders incremental deployment of features (as also identified by~\cite{matls}), which may lead to security degradation and ossification of QUIC.
To address these issues, we will therefore explore explicit feature negotiation mechanisms for SMAQ in order to decouple the requirements.

\afblock{Application Data Security.}
To ensure end-to-end security of application data, XADS provides an additional encryption layer, where currently only \texttt{STREAM} frames are considered.
Hence, XADS needs to be extended to also encrypt other frame types~\cite[Sec. 19]{rfc9000}, as well as other QUIC extensions carrying application data such as the \textit{Unreliable Datagram Extension}~\cite{rfc9221}.

 \section{Related Work}
\label{sec:related-work}

Middleboxes for connection splitting or performance enhancement have a
long history, dating back to at least the mid-1990s when wireless and
satellite links were ``optimized'' for (mobile) Internet usage, some
common practices documented by the IETF~\cite{rfc3135}.  With
virtually ubiquitous TLS and now the uptake of QUIC, application and
transport layer information is no longer accessible to intermediaries,
requiring explicit integration of intermediaries.
Investigationg QUIC and HTTP/3 performance over satellite links, Kosek et al. ~\cite{quic.satcom} showed the benefits of QUIC PEPs for SATCOM; yet, the protection of application data from middlebox access was not considered, rendering the concept unsuitable for practical use.
On the other hand, middlebox-aware TLS (maTLS,~\cite{matls} and multi-context TLS (mcTLS~\cite{mctls}) offer middlebox
support while keeping control of their capabilities, in contrast to
earlier designs that would just split TLS and give full access to application data~\cite{split-tls}.
The MASQUE WG of the IETF explores middlebox
control via HTTP/3 but terminates the controlling QUIC
connection and does not expose protocol state to an intermediary so
that QUIC traffic is only forwarded as opaque packets~\cite{rfc9298}.  Sidecar~\cite{sidecar} is a recent design that uses a side
channel to enable simple operations (prioritizing, delaying, dropping,
etc.) on opaque packets in middleboxes.  \textit{The Onion
Router} (TOR,~\cite{tor}) explicitly expands a connection through TOR relays
hop-by-hop, splitting up the transport connections but achieving
end-to-end security through multiple levels of encryption.
Some early work supported state handover across intermediaries~\cite{i-tcp,mobile-tcp} or servers~\cite{migratory-tcp}, albeit not
yet including security.  QUIC inherently supports connection migration
with the help of the peer and thus can redirect traffic securely~\cite{rfc9000}, which we leverage in this paper.
Conforti et al. showed how QUIC's connection migration can be used for container relocation while preserving ongoing connections~\cite{quic-container-migration}, where the communication continues after the container state is handed over to another machine.

 \section{Conclusion}
\label{sec:conclusion}

In this paper, we enhanced QUIC to selectively expose information to intermediaries, thereby enabling endpoints to consciously insert middleboxes into an end-to-end encrypted QUIC connection while preserving its privacy, integrity, and authenticity.
We evaluated our design in a distributed PEP environment over satellite networks, finding that the performance improvements of SMAQ are dependent on the path and application layer properties: the higher the RTT and loss, and the more data is transferred over a connection, the higher the benefits of SMAQ.
Our findings highlight the potential of SMAQ, warranting further exploration: while advancing the design, problem-spaces such as load balancing or live service migration promise exciting possibilities.

\section*{Acknowledgements}
This work was supported by the Federal Ministry of Education and Research of Germany (BMBF) project 6G-Life (16KISK002).

\vspace{-0.8em}

\bibliographystyle{IEEEtran}
\bibliography{index}

\end{document}